\documentclass[preprintnumbers,amsmath,amssymbm,preprint]{revtex4}
\usepackage{epsfig}
\usepackage{graphicx}
\usepackage{amssymb}

\begin{document}
\title{Self-gravitating ring of matter in orbit around a black hole: The innermost stable circular orbit}
\author{Shahar Hod}
\address{The Ruppin Academic Center, Emeq Hefer 40250, Israel}
\address{ }
\address{The Hadassah Institute, Jerusalem 91010, Israel}
\date{\today}

\begin{abstract}
We study analytically a black-hole-ring system which is composed of
a stationary axisymmetric ring of particles in orbit around a
perturbed Kerr black hole of mass $M$.
In particular, we calculate the shift in the orbital frequency of
the innermost stable circular orbit (ISCO) due to the finite mass
$m$ of the orbiting ring. It is shown that for thin rings of
half-thickness $r\ll M$, the dominant finite-mass correction to the
characteristic ISCO frequency stems from the self-gravitational
potential energy of the ring (a term in the energy budget of the
system which is {\it quadratic} in the mass $m$ of the ring) . This
dominant correction to the ISCO frequency is of order
$O(\mu\ln(M/r))$, where $\mu\equiv m/M$ is the dimensionless mass of
the ring. We show that the ISCO frequency {\it increases} (as
compared to the ISCO frequency of an orbiting test-ring) due to the
finite-mass effects of the self-gravitating ring.
\end{abstract}
\maketitle

\section{Introduction}
The geodesic motions of test particles in black-hole spacetimes are
an important source of information on the structure of the spacetime
geometry \cite{Bar,Chan,Shap,CarC,Fav,Hod1,Hod2,Hod3,Kol}. Of
particular importance is the innermost stable circular orbit (ISCO).
This orbit is defined by the onset of a dynamical instability for
circular geodesics. In particular, the ISCO separates stable
circular orbits from orbits that plunge into the central black hole
\cite{Chan}. This special geodesic therefore plays a central role in
the two-body dynamics of inspiralling compact binaries since it
marks the critical point where the character of the motion sharply
changes \cite{Fav}. In addition, this marginally stable orbit is
usually regarded as the inner edge of accretion disks in composed
black-hole-disk systems \cite{Chan}.

An important physical quantity which characterizes the ISCO is the
orbital angular frequency $\Omega_{\text{isco}}$ as measured by
asymptotic observers. This characteristic frequency is often
regarded as the end-point of the inspiral gravitational templates
\cite{Fav}. For a test-particle in the Schwarzschild black-hole
spacetime, this frequency is given by the well-known relation
\cite{Bar,Chan,Shap,CarC,Fav,Hod1,Hod2,Hod3,Kol}
\begin{equation}\label{Eq1}
M\Omega_{\text{isco}}=6^{-3/2}\  ,
\end{equation}
where $M$ is the mass of the central black hole.

Realistic astrophysical scenarios often involve a composed two-body
system in which the mass $m$ of the orbiting object is smaller but
{\it non}-negligible as compared to the mass $M$ of the central
black hole \cite{Fav}. In these situations the zeroth-order
(test-particle) approximation is no longer valid and one should take
into account the gravitational self-force (GSF) corrections to the
orbit
\cite{Ori,Poi,Lou1,Det1,Bar1,Det2,Sag,Kei,Sha,Dam,Bar2,Fav2,Akc}.
These first-order corrections take into account the {\it finite}
mass $m$ of the orbiting object. The gravitational self-force has
two distinct contributions: (1) It is responsible for dissipative
(radiation-reaction) effects that cause the orbiting particle to
emit gravitational waves. The location of the ISCO may become
blurred due to these non-conservative effects \cite{Ori,Fav}. (2)
The self-force due to the finite mass of the particle is also
responsible for conservative effects which preserve the
characteristic constants of the orbital motion. These conservative
effects produce a non-trivial shift in the ISCO frequency which
characterizes the two-body dynamics.

It should be emphasized that the computation of the conservative
shift in the characteristic ISCO frequency (due to the {\it finite}
mass of the orbiting object) is a highly non-trivial task. A notable
event in the history of the two-body problem in general relativity
took place three years ago: after two decades of intensive efforts
by many groups of researches to evaluate the conservative self-force
corrections to the orbital parameters, Barack and Sago \cite{Bar2}
have succeeded in computing the shift in the ISCO frequency due to
the finite mass of the orbiting object. Their numerical result for
the corrected ISCO frequency can be expressed in the form
\cite{Dam,Bar2}:
\begin{equation}\label{Eq2}
M\Omega_{\text{isco}}=6^{-3/2}(1+c\cdot \mu)\ \ \ \text{with}\ \ \
c\simeq 0.251\ ,
\end{equation}
where
\begin{equation}\label{Eq3}
\mu\equiv m/M\ll 1\
\end{equation}
is the dimensionless ratio between the mass of the orbiting object
(the `particle') and the mass of the black hole. The result
(\ref{Eq2}) provides valuable information about the conservative
dynamics of the composed two-body system in the strong-gravity
regime.

It is worth emphasizing that the $O(\mu)$ correction term to the
ISCO frequency [see Eq. (\ref{Eq2})] stems from similar correction
terms that appear in the metric components of the perturbed
black-hole spacetime \cite{Bar2}. These correction terms to the
metric components of the ``bare" Schwarzschild spacetime are also
linear (in the extreme mass ratio regime) in the dimensionless mass
$\mu$ of the orbiting particle \cite{Bar2}.

In the present study we shall analyze a closely related (but
mathematically much simpler) problem: that of a {\it
self-gravitating} thin ring of matter in equatorial orbit around a
central black hole \cite{Will,Noterec}. This composed system, like
the original black-hole-particle system, is characterized by a
perturbative (finite-mass) correction to the ISCO frequency [see Eq.
(\ref{Eq15}) below]. In fact, as we shall show below, the
leading-order shift in the ISCO frequency can be computed {\it
analytically} for this axisymmetric black-hole-ring system.

Before proceeding into details, it is important to emphasize that
there is one important difference between the black-hole-particle
system studied in
\cite{Ori,Poi,Lou1,Det1,Bar1,Det2,Sag,Kei,Sha,Dam,Bar2,Fav2,Akc} and
the black-hole-ring system that we shall study here:
As emphasized in Ref. \cite{Bar2}, the work \cite{Bar2} is
complementary to the analysis of \cite{Ori} in that \cite{Bar2}
considered only conservative GSF corrections and ignored dissipative
(radiation-reaction) GSF effects. It is only then that the ISCO has
a sharp location. It should be emphasized that, for the
black-hole-particle system, the dissipative effects actually
dominate over the conservative ones \cite{Ori,Bar2}. Thus, the
radius and orbital frequency of the ISCO in the black-hole-particle
system are not sharply defined. On the other hand, due to the axial
symmetry of the black-hole-ring system, there is no emission of
gravitational waves and thus there are no dissipative effects in the
black-hole-ring system. The black-hole-ring system is therefore
characterized by purely conservative finite-mass corrections to the
dynamics. Consequently, the radius and orbital frequency of the ISCO
in the black-hole-ring system are sharply defined [see Eqs.
(\ref{Eq9}) and (\ref{Eq15}) below].

It is worth emphasizing that recently \cite{Hod2} we analyzed a
simplified black-hole-ring toy-model. Following the original
analysis of \cite{Will}, in \cite{Hod2} we ignored the {\it
self}-gravitational potential energy of the ring. This self-energy
term represents the inner interactions between the {\it many}
particles that compose the orbiting ring. Since our goal in
\cite{Hod2} was to model the conservative dynamics of the two-body
(black-hole-particle) system (with a {\it single} orbiting
particle), we did not consider in \cite{Hod2} this {\it
many}-particle self-gravitational term. The omission of this
self-gravitational potential energy term has allowed us to focus in
\cite{Hod2} on the frame-dragging effect caused by the orbiting
object. In this respect, the ring considered in \cite{Will,Hod2}
should be regarded as a quasi test ring.

Our main goal in the present study is to analyze the influence of
this {\it self}-gravitational potential energy term on the dynamics
of the ring (as emphasized above, following the original analysis of
\cite{Will}, this self-energy term was ignored in the toy-model
studied in \cite{Hod2}). As we shall show below, this
self-gravitational potential energy of the ring determines the
leading-order correction to the ISCO frequency in the thin-ring
regime.

\section{The black-hole-ring system} We consider a black-hole-ring
system which is composed of a stationary axisymmetric ring of
particles in orbit around a black hole. This system is characterized
by five physical parameters \cite{Will}: The mass $M$ of the black
hole, the angular momentum per unit mass $a$ of the black hole, the
rest mass $m$ of the ring, the proper circumferential radius $R$ of
the ring, and the half-thickness $r$ of the ring. We shall assume
that the ring is thin and weakly self-gravitating in the sense that
\begin{equation}\label{Eq4}
\text{max}(r/m,1)\ll\ln(M/r)\ll M/m\  .
\end{equation}

The total energy (energy-at-infinity) of the ring in the black-hole
spacetime is given by \cite{Bar,Will,Tho,NoterR}
\begin{eqnarray}\label{Eq5}
E(R;M,a,m,r)=m\cdot {{R^{3/2}-2MR^{1/2}\pm
aM^{1/2}}\over{R^{3/4}(R^{3/2}-3MR^{1/2}\pm 2aM^{1/2})^{1/2}}}
-{{m^2}\over{2\pi R}}\ln(R/r)+O(m^2/M)\ ,
\end{eqnarray}
where the upper/lower signs correspond to
co-rotating/counter-rotating orbits, respectively. The first term on
the r.h.s. of (\ref{Eq5}) represents the first-order contribution of
the ring to the total mass of the system \cite{Bar}. In the
Newtonian (large-$R$) limit it becomes $m-M_{\text{ir}}m/2R$, which
can be identified as the rest mass of the ring plus the (negative)
potential energy of the black-hole-ring system plus the rotational
energy of the ring. The second term on the r.h.s. of Eq. (\ref{Eq5})
represents the (second-order) self-gravitational potential energy of
the ring \cite{Tho}.

The sub-leading correction term $O(m^2/M)$ in Eq. (\ref{Eq5})
represents a non-linear contribution to the energy of the ring which
stems from $O(m/M)$ corrections to the metric components of the
``bare" Kerr spacetime \cite{Will}. In the $\ln(M/r)\gg1$ regime
[the thin-ring regime, see Eq. (\ref{Eq4})] this $O(m^2/M)$ term is
much smaller than the self-gravitational potential energy of the
ring which is of order $O\big({{m^2}\over{M}}\ln(R/r)\big)$, see Eq.
(\ref{Eq5}). Thus, the dominant contribution to the ISCO
frequency-shift [a term of order
$O\big({{m}\over{M^2}}\ln(M/r)\big)$, see Eq. (\ref{Eq15}) below]
would stem from the self-gravitational potential energy of the thin
ring. This correction term would dominate over a sub-leading
correction term [of order $O\big({{m}\over{M^2}})$] which stems from
finite-mass corrections to the metric components of the ``bare" Kerr
metric \cite{Will}.

\section{The innermost stable circular orbit} A standard way to identify the location of the
ISCO is by finding the minimum of the orbital energy
\cite{Fav,Fav2,Bon,Noteisc}. A simple differentiation of (\ref{Eq5})
with respect to $R$ yields the characteristic equation
\cite{Notecons,Notedif}
\begin{equation}\label{Eq6}
{{R^{1/4}(R^2-6MR \pm
8aM^{1/2}R^{1/2}-3a^2})\over{(R^{3/2}-3MR^{1/2}\pm
2aM^{1/2})^{3/2}}} +\mu\cdot[\pi^{-1}\ln(R/r)+O(1)]=0\
\end{equation}
for the location of the ISCO. The zeroth-order solution (with
$\mu\equiv 0$) of the characteristic equation (\ref{Eq6}) is given
by \cite{Bar}
\begin{equation}\label{Eq7}
R_0=M\{3+Z_2\mp[(3-Z_1)(3+Z_1+2Z_2)]^{1/2}\}\  ,
\end{equation}
where
\begin{equation}\label{Eq8}
Z_1\equiv 1+(1-a^2/M^2)^{1/3}[(1+a/M)^{1/3}+(1-a/M)^{1/3}]\ \ \
\text{and}\ \ \ Z_2\equiv (3a^2/M^2+Z^2_1)^{1/2}\  .
\end{equation}
The corrected (first-order) solution of the characteristic equation
(\ref{Eq6}) is then given by
\begin{equation}\label{Eq9}
R_{\text{isco}}=R_0\big\{1-\mu\cdot f(a)\cdot[\ln(M/r)+O(1)]\big\}\
,
\end{equation}
where
\begin{equation}\label{Eq10}
f(a)\equiv
{{1}\over{2\pi}}\Big(1-{{3M}\over{R_0}}\pm{{2aM^{1/2}}\over{R^{3/2}_0}}\Big)^{1/2}\
.
\end{equation}

The expression (\ref{Eq9}) provides the location of the ISCO for the
composed black-hole-ring system in the finite mass-ratio
(finite-$\mu$) regime. In the Schwarzschild ($a\to0$) limit one
finds
\begin{equation}\label{Eq11}
R_{\text{isco}}(a=0)=6M\big[1-\mu\cdot{{1}\over{2\sqrt{2}\pi}}\ln(M/r)\big]\
.
\end{equation}
In the extremal ($a\to M$) limit one finds
\begin{equation}\label{Eq12}
R_{\text{isco}}(a=M)=9M\big[1-\mu\cdot{{2}\over{3\sqrt{3}\pi}}\ln(M/r)\big]\
.
\end{equation}
for counter-rotating orbits, and
\begin{equation}\label{Eq13}
R_{\text{isco}}(a=M(1-\epsilon))=M\big\{1+(4\epsilon)^{1/3}\big[1-\mu\cdot
{{\sqrt{3}}\over{4\pi}}\ln(M/r)\big]\big\}\ .
\end{equation}
for co-rotating orbits, where $\epsilon\ll1$.

The angular velocity of the ring is given by \cite{Chan,Will}
\begin{equation}\label{Eq14}
\Omega={{\sqrt{{{M}/{R^3}}}}\over{\pm1
+a\sqrt{{{M}/{R^3}}}}}[1+O(\mu)]\ .
\end{equation}
The correction term $O(\mu)$ in Eq. (\ref{Eq14}) represents a
non-linear contribution to the angular velocity of the ring
\cite{Will} which stems from $O(\mu)$ corrections to the metric
components of the ``bare" Kerr spacetime. In the $\ln(M/r)\gg1$
regime [the thin-ring regime, see Eq. (\ref{Eq4})] this $O(\mu)$
term is much smaller than the leading-order $O(\mu\ln(M/r))$
correction term to $\Omega_{\text{isco}}$ which stems from the
self-gravitational potential energy correction to $R_{\text{isco}}$
[a term of order $O(\mu\ln(M/r))$, see Eq. (\ref{Eq9})].

Substituting (\ref{Eq9}) into (\ref{Eq14}), one finds
\begin{equation}\label{Eq15}
\Omega_{\text{isco}}=\Omega_0\big\{1+\mu\cdot
g(a)\cdot[\ln(M/r)+O(1)]\big\}
\end{equation}
for the perturbed ISCO frequency of the ring, where
\begin{equation}\label{Eq16}
\Omega_0\equiv{{\sqrt{{{M}/{R^3_0}}}}\over{\pm1
+a\sqrt{{{M}/{R^3_0}}}}}\
\end{equation}
is the zeroth-order frequency of an orbiting test-ring, and
\begin{equation}\label{Eq17}
g(a)\equiv
{{3}\over{4\pi}}(1-a\Omega_0)\Big(1-{{3M}\over{R_0}}\pm{{2aM^{1/2}}\over{R^{3/2}_0}}\Big)^{1/2}\
.
\end{equation}

The expression (\ref{Eq15}) provides the characteristic frequency of
the ISCO for the composed black-hole-ring system in the finite
mass-ratio (finite-$\mu$) regime. In the Schwarzschild ($a\to 0$)
limit one finds
\begin{equation}\label{Eq18}
M\Omega_{\text{isco}}(a=0)=6^{-3/2}\big[1+\mu\cdot{{3}\over{4\sqrt{2}\pi}}\ln(M/r)\big]\
.
\end{equation}
[Compare (\ref{Eq18}) with the corresponding result (\ref{Eq2}) for
the (Schwarzschild-)black-hole-particle system. In both cases the
ISCO frequency increases due to the {\it finite} mass of the
orbiting object. Note, however, that the $O(\mu)$ correction in Eq.
(\ref{Eq2}) stems from $O(\mu)$ corrections to the metric components
of the bare Schwarzschild metric while the dominant $O(\mu\ln(M/r))$
correction in Eq. (\ref{Eq18}) stems from the self-gravitational
potential energy of the orbiting ring]. In the extremal ($a\to M$)
limit one finds
\begin{equation}\label{Eq19}
M\Omega_{\text{isco}}(a=M)=-{1\over
26}\big[1+\mu\cdot{{9\sqrt{3}}\over{26\pi}}\ln(M/r)\big]\
\end{equation}
for counter-rotating orbits, and
\begin{equation}\label{Eq20}
M\Omega_{\text{isco}}(a=M(1-\epsilon))={1\over2}\big\{1-{3\over4}(4\epsilon)^{1/3}
[1-\mu\cdot{{\sqrt{3}}\over{4\pi}}\ln(M/r)\big]\big\}
\end{equation}
for co-rotating orbits.

It is worth emphasizing that the shift-function $g(a)$ is a
non-negative function for all $a$-values. Thus, taking cognizance of
Eq. (\ref{Eq15}) one concludes that the ISCO frequency {\it
increases} (in its absolute value) due to the finite-mass effects of
the orbiting ring:
\begin{equation}\label{Eq21}
|\Omega_{\text{isco}}|>|\Omega_0|\  .
\end{equation}

\section{Summary and Discussion} We have analyzed a stationary and axisymmetric
black-hole-ring system which is composed of a {\it self-gravitating}
ring of matter in orbit around a central black hole. In particular,
we have calculated the shift in the fundamental frequency of the
innermost stable circular orbit (ISCO) due to the finite mass of the
ring. For thin rings with $\ln(M/r)\gg \text{max}(r/m,1)$ [see Eq.
(\ref{Eq4})], the dominant finite-mass correction to the ISCO
frequency stems from the self-gravitational potential energy of the
ring. This correction term is of order $O[\mu\ln(M/r)]$ and, in the
thin-ring regime (\ref{Eq4}), it dominates over a sub-leading
correction term of order $O(\mu)$ which stems from $O(\mu)$
corrections to the metric components of the ``bare" black hole. We
have shown [see Eqs. (\ref{Eq15}) and (\ref{Eq17})] that the
characteristic ISCO frequency {\it increases} due to the finite-mass
effects of the self-gravitating orbiting ring.

It is worth emphasizing that the composed black-hole-ring system,
being axially symmetric, is characterized by purely conservative
gravitational effects. That is, there is no emission of
gravitational waves in this axially symmetric system. Thus, this
composed two-body system probably has a limited observational
relevance. The black-hole-ring system should instead be regarded as
a simple toy-model for the astrophysically more relevant two-body
(black-hole-particle) system.

In this respect, the black-hole-self-gravitating-ring model has two
important advantages over the astrophysically more realistic
black-hole-particle system:
\newline
(1) The original black-hole-particle system is a highly
non-symmetrical system. This lack of symmetry makes the computation
of the ISCO frequency shift a highly non-trivial task. In fact, one
is forced to use numerical techniques \cite{Bar2} in order to
compute the ISCO shift in this system [see Eq. (\ref{Eq2})]. On the
other hand, the black-hole-ring toy model is an axially symmetric
system. As we have shown above, this axial symmetry of the
black-hole-ring model simplifies the calculation of the ISCO
frequency shift. In fact, we have seen that, due to the axial
symmetry of the black-hole-ring system, one can obtain an {\it
analytic} formula [see Eq. (\ref{Eq15})] for the ISCO frequency
shift in this composed system. We believe that any new analytical
solution, even one for a simplified (more symmetrical) problem, is
certainly a useful contribution to this field.
\newline
(2) To the best of our knowledge, the highly important result
(\ref{Eq2}) for the ISCO frequency shift in the
Schwarzschild-black-hole-particle system has so far not been
extended to the case of rotating Kerr black holes. This lack of
results for generic Kerr black holes is probably due to the
numerical complexity of the problem. On the other hand, as we have
seen above [see Eq. (\ref{Eq15})], the calculation of the ISCO
frequency shift in the composed black-hole-ring model can be
extended to the regime of {\it generic} (that is, rotating) Kerr
black holes. In this respect, it is worth noting that our conclusion
(\ref{Eq18}) that the ISCO frequency of the
Schwarzschild-black-hole-ring system increases due to the finite
mass of the orbiting object is in agreement with the corresponding
result (\ref{Eq2}) for the original
Schwarzschild-black-hole-particle system \cite{Bar2}. This
qualitative agreement may indicate that the increase in the ISCO
frequency (due to the finite mass of the orbiting object) may be a
generic feature of the conservative two-body dynamics [see Eq.
(\ref{Eq21})].

Finally, it is worth mentioning the well known fact that many
astrophysical black holes have accretion disks around them
\cite{Lemos}. The radial location of the test particle ISCO [see Eq.
(\ref{Eq7})] is usually regarded as the inner edge of the accretion
disk in these composed black-hole-disk systems. It is expected,
however, that self-gravitational effects (due to the finite mass of
the accretion disk) would modify the location (the radius) of the
disk's inner edge in these astrophysical black-hole-disk systems. In
this respect, our result (\ref{Eq9}) for the location of the ISCO in
the composed black-hole-self-gravitating-ring system should be
regarded as a first approximation for the location of the ISCO (the
location of the disk's inner edge) in realistic black-hole-disk
systems. We believe that our analytic treatment of the
black-hole-ring system will be useful and stimulating for further
studies of the physical properties (and, in particular, the location
of the ISCO) of astrophysical black-hole-disk systems.

\bigskip
\noindent
{\bf ACKNOWLEDGMENTS}
\bigskip

This research is supported by the Carmel Science Foundation. I thank
Yael Oren, Arbel M. Ongo and Ayelet B. Lata for helpful discussions.



\begin{thebibliography}{99}
\bibitem{NoterR} The first term on the r.h.s of Eq. (\ref{Eq5}) should be multiplied by a correction
factor $1+O(r/R)$ which stems from the finite thickness of the ring.
Note that in the thin-ring regime, $r/m\ll \ln(R/r)$ [see Eq.
(\ref{Eq4})], this correction term is dominated over by the
self-gravitational potential energy of the thin-ring which is of
order $O(m^2\ln(R/r)/R)$ [see the second term on the r.h.s of Eq.
(\ref{Eq5})].
\bibitem{Noteisc} The critical orbit so defined is sometimes referred to as the MECO
(maximum binding-energy circular orbit) \cite{Fav,Fav2,Bon}.
\bibitem{Notecons} We consider here a family of orbiting rings with a fixed (constant)
value of the parameter $r$, the half-thickness of the rings.
Alternatively, one may consider a family of constant proper-density
rings. In this case, the parameter $r$ would have the dependence
$r=const\times R^{-1/2}$, where $R$ is the circumferential radius of
the ring. For such family of constant proper-density rings, the
square brackets in Eq. (\ref{Eq6}) would simply acquire a
sub-leading correction term of order $O(1)$ which, in the thin-ring
regime $r\ll R$ [see Eq. (\ref{Eq4})], is dominated over by the
leading-order correction term $\ln(R/r)\gg1$ [see Eq. (\ref{Eq6})].
\bibitem{Notedif} The first term on the l.h.s of Eq. (\ref{Eq6}) should be multiplied by a
correction factor $1+O(r/R)$ with $r/R\ll1$ [see Eq. (\ref{Eq4})].
\end{thebibliography}
\end{document}